\documentclass{aa}
\usepackage{graphics}

\def\be{\begin{equation}}
\def\en{\end{equation}}

\begin{document}

\thesaurus{02(02.07.2; 03.13.4; 12.03.4; 12.12.1)}

\title{Gravitational waves from galaxy cluster distributions}

\author{Vicent Quilis\inst{1}
\thanks{vicent.quilis@durham.ac.uk}
\and Jos\'e M$^{\underline{\mbox{a}}}$. Ib\'a\~nez\inst{2}
\thanks{jose.m.ibanez@uv.es}
\and Diego S\'aez\inst{2} 
\thanks{diego.saez@uv.es}}

\institute{Department of Physics, University of Durham, South Road, 
DH1 3LE, Durham, UK
\and 
Departament d'Astronomia i
Astrof\'{\i}sica, Universitat de Val\`encia,
E-46100 Burjassot, Val\`encia, Spain}

\offprints{V. Quilis}

\maketitle

\begin{abstract}

Galaxy clusters are sources of gravitational radiation. The main 
aim of this paper is to give numerical estimates and theoretical 
description of the relevant features of the gravitational radiation
coming from an appropriate spatial distributions of galaxy clusters. 
Since no analytical approaches are currently available to describe 
the strongly nonlinear regime, our numerical approach --combining 
numerical simulations with statistical arguments -- seems to be an 
useful way of studying the main features of that radiation. 
Although far to be detectable with present technology, we advance some
ideas about future observational strategies and its cosmological 
implications. 

\keywords{cosmology:theory -- gravitational waves -- 
large-scale structure of universe -- methods: numerical}

\end{abstract}

\section{Introduction}
The cosmological stochastic gravitational-wave background produced 
by mildly nonlinear evolution of density fluctuations can be estimated 
using analytical perturbative methods (see, e.g., Matarrese \& Mollerach 
1997 and  
references cited therein). However, the estimation of that emission, 
in the strongly nonlinear regime, would require non-perturbative 
approaches which are not currently available. Galaxy clusters are 
evolving beyond the mildly nonlinear regime and they undergo the 
so-called virialization process (violent relaxation). Then, the 
following question arises (Matarrese \& Mollerach 1997): have the clusters 
produced a significant
amount of gravitational radiation? Since the problems with the 
{\em analytical} calculation of the gravitational radiation
coming from strongly nonlinear galaxy clusters has not been still 
solved, we have estimated that radiation numerically. 

In a previous paper (Quilis et al. 1998b), the gravitational radiation 
released during the formation of an isolated galaxy cluster
was analyzed.
In this paper, we have extended that calculation to the case of the 
emission coming from spatial distributions of galaxy clusters, thus 
we have performed, for the first time, a quantitative nonlinear 
calculation of this 
emission. Hence, we have obtained a numerical estimate of the
contribution to the stochastic gravitational-wave background 
due to a given spatial distribution of galaxy clusters.
Although the resulting emission has appeared to be weak in our
calculations, we want to stress that, at least from the theoretical 
point of view, it is worthy to know the main features of that 
radiation.

Cluster simulations including both baryonic and dark matter have undergone 
a high level of development (e.g. Kang et al. 1994, Metzler \& Evrard 1994, 
Navarro et al. 1995, Anninos \& Norman 1996, Gheller et al. 1998). The 
baryonic component is evolved either 
with an Eulerian or with a Lagrangian code, the treatment of the dark matter 
component is based on appropriate N-body techniques, and both components are 
gravitationally coupled through Poisson's equation, which is usually solved 
using the Fast Fourier Transform. Simulations based on a 3D Eulerian code 
(Quilis et al. 1998a) 
were used to estimate the gravitational radiation generated by individual 
galaxy clusters (Quilis et al. 1998b). Computations were performed in the 
framework of the
standard Cold Dark Matter (CDM) scenario. The Hubble constant and the density 
parameter were assumed to be $H_{0}=50 \ {\rm Km s}^{-1} {\rm Mpc}^{-1}$ and 
$\Omega_{0} =1$, respectively. The same assumptions and method have been used 
in the cluster simulations used in this paper (this procedure is justified 
below). Appropriate initial conditions have been chosen to get both rich 
and standard galaxy clusters. Our rich clusters have a X-ray luminosity 
($L_{x}$) of the order of $\sim 10^{44} \ {\rm erg/s}$, a temperature 
$T \geq 3 \times 10^{7} \ K$ and, a total mass inside the Abell radius 
($3 \ {\rm Mpc}$) $M \geq 4 \times 
10^{14} \ M_{\odot}$. Our standard cluster has 
$L_{x} \sim 10^{43} \ {\rm erg/s}$, 
$T \leq 6 \times 10^{6}$ and, $M \leq 10^{14} 
\ M_{\odot}$.

As it was discussed in Quilis et al. (1998b), the gravitational radiation 
from a galaxy cluster
produces progressive deformations on some material systems, this is what was 
referred to as {\em the secular effect}. In the particular case of a system 
formed by two test particles, the deformation reduces to a relative variation 
of their separation distance. Since this variation is proportional to time,
then, leaving aside the question on the characteristic frequency (crucial for
detectability), it reaches values which are in the range of space-based laser 
interferometric observatories of gravitational waves.

There are various conclusions of Quilis et al. (1998b) to be taking into 
account here: 
(i) The secular effect is approximately proportional to $D^{-1}$, where $D$ 
is the distance from the cluster to the observer, (ii) clusters located at a 
distance $D>600 \ {\rm Mpc}$ only produce 
a small effect which is neglected in the 
simulations of this paper, (iii) for distances $D<600 \ {\rm Mpc}$, 
gravitational 
waves can be considered as propagating in the Minkowskian space tangent to 
the Friedman-Robertson-Walker spacetime at the emission point. 
Although the secular effect is only significant in the near zone ($D<600 
\ {\rm Mpc}$), no 
differentiation is made -- along the paper -- between this effect 
and gravitational waves (in the radiation zone), this is because, 
in both cases, the 
deformations produced on detectors have the same sources (galaxy clusters) 
and the same transverse signature and, also, because these deformations are 
estimated with the same formulae holding in both the near and the radiation 
zones, (iv) the secular effect produced by each cluster does not appear
isolated, but superimposed to the effect due to other clusters, and (v) the 
internal dynamics of clusters having similar features ($L_{x}$, $T$, and $M$) 
is expected to depend on the cosmological parameters ($\Omega_{0}$, $H_{0}$,...)
weakly, while the spatial distributions of clusters is sensitive to these 
parameters. Consequently with this last point, we have fixed $\Omega_{0}$ and 
$H_{0}$ for simulating individual clusters, while the chosen spatial 
distributions mimic some features of cluster catalogues. 

\section{Simulation strategy and results}

Only very crude simulations of the total secular effect produced by cluster 
distributions were given in Quilis et al. (1998b), where some numbers were 
found with the 
essential aim of proving that the effects of many clusters do not cancel 
among them. 
The main goal of this paper is, exactly, this one: to present improved 
calculations showing the features of the signal coming from an appropriate
spatial distribution of galaxy clusters. This spatial distributions of clusters 
could be chosen either mimicking the observed one or simulating them in a 
certain scenario of structure formation in the Universe. In any case, the 
distribution can 
be restricted inside a sphere of $600 \ {\rm Mpc}$ radius. According 
to Dalton et al. (1994), the analysis of the APM galaxy survey leads to a 
mean 
density of rich clusters of $4.25 \times 10^{-6} \ 
{\rm Mpc}^{-3}$ and a two-point 
correlation function of the form $\xi_{cc}=( r_{0} / r )^{2}$, with 
$r_{0}=28.6 \ {\rm Mpc}$. All the 
distributions of rich galaxy clusters used in this 
paper -- to estimate the total secular effect -- have been constrained 
(Pons-Borderia et al. 1999) to 
have these features.

In order to give a complete and clear physical description of the secular
effect, a simple detector formed by two test particles, A and B, is 
appropriate. 
Let us analyze the response of this detector to the gravitational waves from 
a cluster (secular effect). In the Transverse Traceless (TT) gauge, the 
relative motion of these particles is fully given (Misner et al. 1973) by 
the quantities 
$h^{^{TT}}_{ij}$, which describe a propagating small perturbations of the 
space-time structure (gravitational wave). Direction $x^{3} \equiv z$ coincides 
with the line of sight of the cluster producing the gravitational radiation, 
while directions $x^{1} \equiv x$ and $x^{2} \equiv y$ are in a plane orthogonal
to this line.  There are four nonvanishing components of $h^{^{TT}}_{ij}$ 
satisfying the relations $h^{^{TT}}_{xy} = h^{^{TT}}_{yx}$ and $h^{^{TT}}_{yy} 
= - h^{^{TT}}_{xx}$; hence, only $h^{^{TT}}_{xx} \equiv h_{+}$ and 
$h^{^{TT}}_{xy} \equiv h_{\times}$ are independent quantities defining two 
polarization states. In TT gauge, there is a system of coordinates attached to 
A, in which the coordinate of the particle B undergoes the following variations 
from present time $t_{0}$ to time $t_{0} + \Delta t$:
\be
\Delta x = \frac {1} {2} \{ x_{0} \dot {h}_{+}(t_{0}) + 
y_{0} \dot {h}_{\times}(t_{0}) \} \Delta (t)
\label{detec1}
\en 
\be
\Delta y = \frac {1} {2} \{ x_{0} \dot {h}_{\times}(t_{0}) - 
y_{0} \dot {h}_{+}(t_{0}) \} \Delta (t)
\label{detec2}
\en 
\be
\Delta z = 0
\label{detec3}
\en 
where $\Delta t$ is much smaller than the period of the gravitational waves 
emitted by clusters ($\Delta t << 10^{9} \ {\rm yr}$), the subscript 
"$0$" stands for 
the initial coordinates of the particle $B$ at time $t_{0}$, the overdot 
stands for a time derivative and, the quantities $\dot {h}_{+}$ and 
$\dot {h}_{\times}$ are computed at point $A$ and at present time.

Each cluster produces a secular effect described by Eqs. (\ref{detec1}) -- 
(\ref{detec3}). These equations show that the effect is proportional to 
$\Delta t$. The estimate of $\Delta x$ and $\Delta y$ requires the knowledge 
of $\dot {h}_{+}$ and $\dot {h}_{\times}$ for the chosen cluster. An explicit 
computation of these quantities is only possible if cluster evolution is known, 
as it occurs in the case of numerically simulated clusters. The total secular 
effect produced by a distribution of clusters can be obtained from the 
$\dot {h}_{+}$ and $\dot {h}_{\times}$ quantities corresponding to each cluster.
Obviously, the simulation of each one of the $\sim 4000$ rich clusters located
inside a sphere of $600 \ {\rm Mpc}$ radius is out of current computational 
capabilities; therefore, some type of statistical treatment of the problem is 
necessary. Quantities $\dot {h}_{+}$ and $\dot {h}_{\times}$ must be assigned 
to each one of the clusters belonging to some spatial distribution using 
adequate criteria. Let us motivate these criteria listing various 
considerations:

(a) The gravitational radiation from a cluster is the superposition of the 
radiation produced by the motion of many particles of dark and baryonic matter 
inside the cluster; hence, low levels of polarization are expected and the 
assumption that quantities $\dot {h}_{+}$ and $\dot {h}_{\times}$ are 
independent is good enough. Furthermore, clusters radiate incoherently and, 
consequently, the waves from different clusters have distinct uncorrelated 
phases. At emission time, each cluster should be in an evolution state 
different from and independent on the state of any other cluster.

(b) Several simulations of rich clusters have been analyzed in order to define 
an interval where the values of $\dot {h}_{+}$ and $\dot {h}_{\times}$ are 
distributed. Let us assume a sphere of $100 \ {\rm Mpc}$ 
radius centered in one of
our simulated clusters. We could observe the central cluster from any point on 
the sphere. The radius passing through the observation point would be our $z$ 
axis and, taking into account the evolution law of the cluster -- which has 
been numerically simulated --, quantities  $\dot {h}_{+}$ and 
$\dot {h}_{\times}$ can be computed for each observation point. Then, a great 
number of these points can be considered and the maximum and minimum of 
these quantities can be easily estimated. The mean of the maxima obtained 
from all the simulated rich clusters, $\bar{\dot {h}} \sim \langle 
\dot {h}_{+}(max) \rangle \sim \langle \dot {h}_{\times}(max) \rangle $, has 
appeared to be $\sim 9. \times 10^{-21} \ {\rm yr^{-1}}$; hence, we 
assume that quantities 
$\dot {h}_{+}$ and $\dot {h}_{\times}$ are distributed in the interval 
($-\bar{\dot {h}},\bar{\dot {h}}$). 

(c) The same study has been done for standard clusters. In this case, the 
resulting $\bar{\dot {h}}$ value is 
$ \sim 1. \times 10^{-21} \ {\rm yr^{-1}}$; hence, rich 
clusters produces an effect which is one order of magnitude greater than that 
of the standard clusters. On account of this fact, only the distributions of 
rich clusters are considered in this paper; nevertheless, more work should be 
done to estimate the contribution of standard and small clusters, which are 
abundant structures producing weak secular effects.

After these comments, the following method seems to be appropriate to give 
values to $\dot {h}_{+}$ and $\dot {h}_{\times}$: According to (a), these 
quantities are generated as statistically independent numbers and, on account 
of (a) and (b), each of these quantities is assumed to be a random number
($\eta $ for $h_{+} $ and $\xi $ for $h_{\times}$) uniformly distributed in 
the interval (-1,1) multiplied by the mean value $\bar{\dot {h}}$. Finally, 
if a given cluster is not located at $100 \ {\rm Mpc}$ from 
the observer, but at a 
distance $D$, number $\bar{\dot {h}}$ must be multiplied by the factor $100/D$.
These criteria plus Eqs. (\ref{detec1}) -- (\ref{detec3}) allow us to compute 
the relative variation of the distance AB produced by an arbitrary cluster 
located at distance $D$ (in Mpc) from the detector. After trivial algebra, the 
following key equation is easily found: 
\be
\frac { \Delta l}{l} = \pm \frac {50}{D} \dot {\bar{h}} (\eta^{2} 
+ \xi^{2})^{1/2} (sin^{2} \theta) \Delta t
\label{key2}
\en  
where $\theta$ is the angle formed by the segment AB and the cluster line of 
sight. For $\theta=0$, the segment AB is aligned with the line of sight and 
no deformation is produced at all. In the case $\theta = \pi /2$, the segment 
AB is orthogonal to the line of sight and the cluster produces a maximum 
$\displaystyle{\frac { \Delta l}{l}}$ independent on the orientation of the 
AB segment in the plane orthogonal to the line of sight.

Given a spatial distribution of clusters and an orientation of the segment 
AB, Eq. (\ref{key2}) allows us to assign a small $\displaystyle{\frac 
{ \Delta l}{l}}$ to each cluster and, then, all these values must be added 
to find the total secular effect produced by the cluster distribution. If 
this distribution is not altered, but the orientation of the segment AB is 
changed, the total secular effect measured by the detector changes 
({\em anisotropy}). According to Eq. (\ref{key2}), this change occurs because
numbers $\eta$, $\xi$, and  $\dot {\bar{h}}$ keep unaltered, but the angle 
$\theta$ corresponding to each cluster varies. Given a spatial distribution of
rich clusters, we have all the ingredients necessary to generate full maps of 
the sky; namely, maps where all the AB orientations are considered. These maps 
completely describe the anisotropy of the total secular effect.

One map of the full sky is displayed in Fig. 1, where quantity $\mu = l^{-1} 
\frac {\Delta l} {\Delta t}$ is given for all the directions joining the center 
of a sphere (A) with the points (B) located on a hemisphere. There are 
anisotropies in the sense that, for the chosen cluster distribution, the $\mu$ 
value --displayed by the grey scale-- depends on the orientation of the segment 
AB in the space. The map of this Figure 
has been expanded in spherical harmonics 
turning out in the superposition of a 
monopole $a_{0,0}=-8.17\times10^{-20}$ and 
the following independent quadrupole components: the real component
$a_{2,0}=-1.38\times10^{-20}$, and the complex ones $a_{2,2}=a^{*}_{2,-2}=
(1.44\times10^{-20}, -1.78\times10^{-20})$ and $a_{2,1}=-a^{*}_{2,-1}=(-4.25
\times10^{-20}, 1.81\times10^{-20})$. Any other multipole appears to be 
negligible. These numerical results are easily understood taking into account
Eq. (\ref{key2}), where the dependence on $sin^{2} \theta$ shows that the 
effect of each cluster is the superposition of a monopole and a quadrupole. 
We can conclude that the total secular effect can be completely described 
by the six components of the quantities ($a_{0,0}$, $a_{2,2}$, $a_{2,1}$, 
$a_{2,0}$), which depend on the chosen cluster distribution.

\begin{figure*}
\resizebox{12cm}{!}{\includegraphics{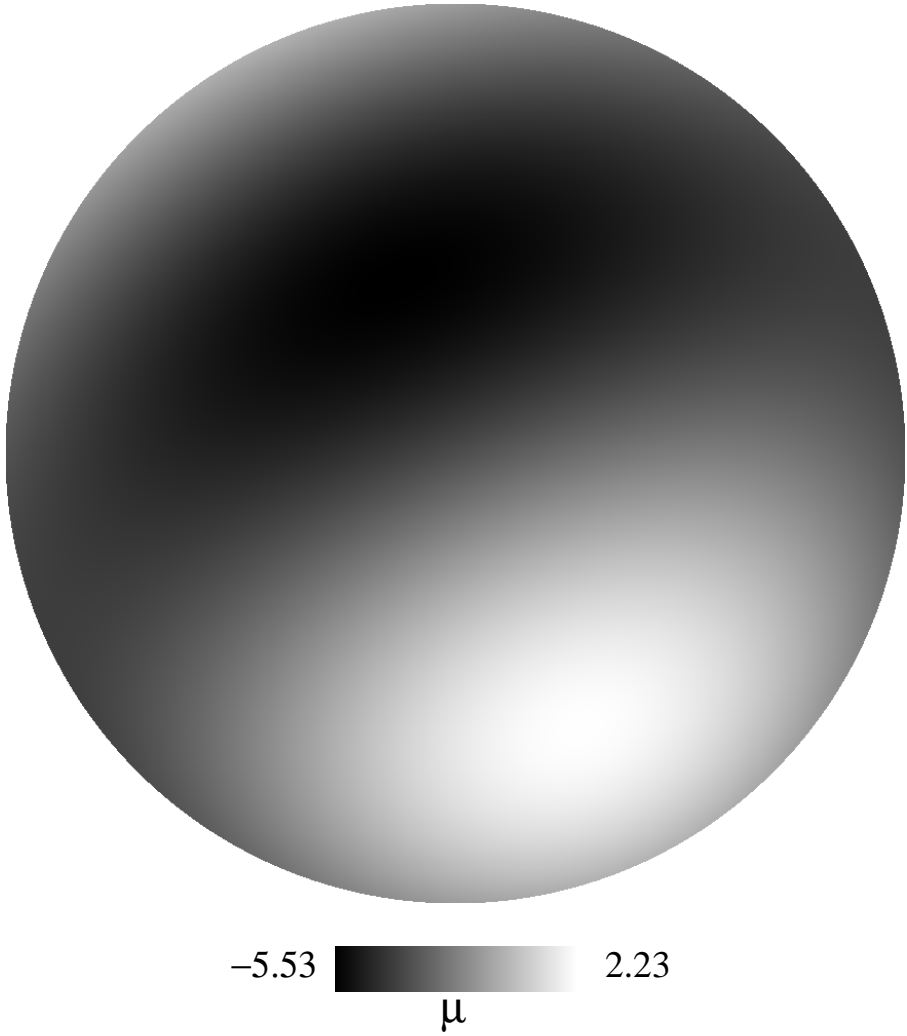}}
\hfill
\parbox[b]{55mm}{
\caption{Map of the secular effect produced by a  
distribution of rich clusters. The quantity $\mu\times10^{20}$, with
$\mu=l^{-1}\frac{\Delta l}
{\Delta t}$ in units of ${\rm yr}^{-1}$, is plotted for a
semisphere. Lambert's projection is used.}}
\end{figure*}

Various simulations have been done using different distributions of clusters. 
Many orientations of the segment AB have been considered, and the mean and 
standard deviation $\sigma $ of the predicted $\mu$ values have been calculated. 
The mean value of $\mu$ changes from simulation to simulation and the standard 
deviation is a rather stable quantity. This quantity, which measures typical 
deviations with respect to the mean and, consequently, anisotropy, ranges in
the interval ($1. \times 10^{-20} \ 
{\rm yr}^{-1}, 2. \times 10^{-20} \ {\rm yr}^{-1}$). 
In the worst case, $\frac {\Delta l} {l} = \mu \Delta t$ takes on the values 
$\sim 10^{-26}$, $\sim 10^{-22}$ and $\sim 10^{-19}$ in time intervals of 
$30 \ {\rm s}$, $3.6$ days and a decade, respectively. 

\section{Discussion and speculations on observational strategies}

The efficiency of the gravitational wave emission from a rich cluster is 
very low, $\approx 10^{-16}$ (see Quilis et al. 1998b). This means that 
during all the age 
of the Universe a single cluster would radiate a gravitational energy of 
$\approx 10^{-16}$ times the cluster mass. The contribution of that 
energy to the present density parameter, $\Omega_{gw}$, would be of 
the order of $\approx 10^{-16}$ $\Omega_c$, being  $\Omega_c$ the 
contribution of all the clusters to the density parameter. That rough 
argument suggests that $\Omega_{gw}$ would range between $10^{-17}$ -- 
$10^{-18}$. A very low value.

The gravitational waves from  galaxy clusters produce anisotropies 
on the Cosmic Microwave Background (CMB). Other gravitational waves  
generated during inflation or in another early process would lead
to "primary" CMB anisotropies because they were
present at recombination time; however, the gravitational waves 
coming from galaxy clusters were emitted after recombination 
and they can only produce "secondary" anisotropies, which are due
to the motion of the CMB photons in the time varying gravitational
field associated to these waves. Although these anisotropies exist
(detailed estimates are in progress),   
they are expected to be too small --for detection-- 
due to the low values of $\Omega_{gw}$ given above.
This means that 
observations of the CMB seem not to be appropriate 
for detecting the gravitational 
background from galaxy clusters. 
Could we use interferometry, as in standard 
detectors of gravitational waves, to detect this background? 

\subsection{Some clues for future observational strategies} 

Measurements of the total secular effect do not require a fixed orientation 
of the AB segment; suppose, for instance, that particles A and B move with the 
terrestrial equator; then, the AB direction depends on time in a well known way 
and it covers all the directions of the equatorial plane during a day.
Every day, the total relative variation $\Delta l / l$ is given by the integral
\be
(\Delta l / l)_{{\rm day}} = \frac {1} {W} \int_{0}^{2 \pi} \mu (\beta) 
 d \beta
\label{equa}
\en 
where angle $\beta$ defines the AB direction inside the equatorial plane
and $W$ is Earth's angular velocity. We have fixed a cluster distribution and, 
then, various planes playing the role of the equatorial one have been 
considered. See, for instance, Fig. 2, where functions $\mu (\beta )$ are 
displayed for three of these planes. Using a large number of planes and 
Eq. (\ref{equa}), many possible values 
of $(\Delta l / l)_{{\rm day}}$ have been 
calculated.  The mean of these values and the standard deviation with respect 
to it are $4.93 \times 10^{-23}$ and $1.8 \times 10^{-23}$.
Other cluster distributions have been considered and the means 
and standard deviations appear to be different; nevertheless, the orders of 
magnitude keep unaltered.
We can conclude that the detector would undergo successive pulses 
with amplitudes of the order of $10^{-23}$ and periods of $6$  hours
(the period is $T= \pi / 2W$, see Fig. 2). The pulse appears 
as a result of the detector rotation in the anisotropic gravitational 
field created by the clusters. 

\begin{figure}
\resizebox{\hsize}{!}{\includegraphics{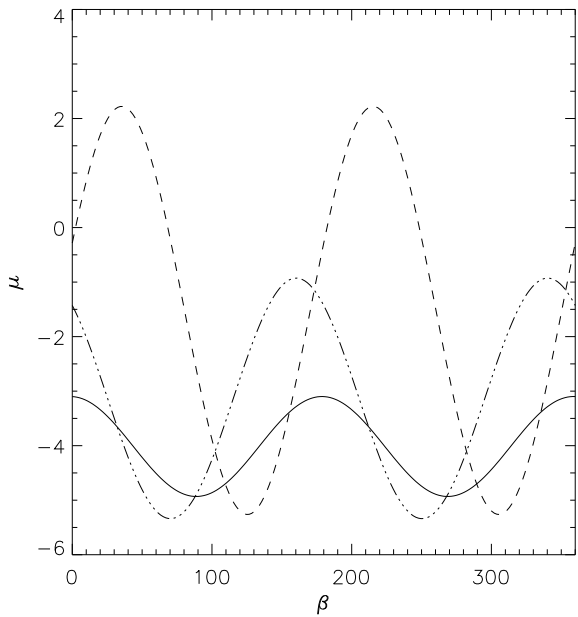}}
\caption{Plot of the quantity $\mu\times10^{20}$ 
(in ${\rm yr}^{-1}$) as a
function of the angle $\beta$ (in degrees) defining the
orientation of the AB segment inside a fixed plane.
Three different planes have been chosen at random.}
\end{figure}

Since the sensitivity of laser interferometry-based detectors decreases 
with the signal frequency $\nu$, they cannot observe --directly--
secular effects with almost negligible $\nu$ ; nevertheless, 
as discussed above, {\em rotating} detectors 
would receive effective pulses -- with frequency $\nu = 2W/ \pi$ --
whose possible detectability would  require further analysis.
The angular velocity of the detector fixes the pulse frequency
$2W/ \pi $ and its amplitude (which is proportional 
to $W^{-1}$ according to Eq. (\ref{equa})). 
According to our estimates,
the amplitude reaches the order $10^{-24}$
for pulse frequencies of the order of $10^{-3}$. 
As the
angular velocity increases, the pulse frequency increases
-- gaining in sensitivity --
but the pulse amplitude decreases. 
Taking into
account these considerations and the fact that interferometry 
is a very accurate technique, the design of specific experiments 
for detecting 
the above systematic and repetitive pulses 
--produced by the clusters on rotating detectors--
deserves attention.

\subsection{Cosmological consequences}

The secular effect and its anisotropy are expected to be dependent 
on the value of some cosmological parameters (see above). Distinct values of 
these parameters would lead to different spatial distributions of clusters and, 
then, to distinct secular signals.
For instance, the dependence of the secular effect on the 
Hubble constant can be easily analyzed. As this constant changes, all the 
distances are multiplied by the factor $H_{0}^{-1}$, and the angles ($\theta$)
between the AB segment and the cluster lines of sight do not change; hence, the 
secular effect, which is roughly proportional to $D^{-1}$, is approximately 
proportional to $H_{0}$.
The secular signals could be predicted for a wide 
range of values of the involved parameters and, then, comparisons of the 
resulting predictions and future observations could give new bounds on some of 
these parameters. These bounds would be very interesting
from the cosmological point of view; however, 
taking into account that: (i)
the secular effect is small and, (ii) its detection 
is not expected to be easy, 
our considerations about the cosmological consequences of detecting 
the gravitational radiation from clusters
have to be read wisely. Other observations could be much more effective
to constraint the cosmological parameters; for instance, CMB
observations from future spatial missions as MAP and PLANCK.

\begin{acknowledgements}
This work has been supported by the Spanish DGES (grants PB96-0797 and 
PB97-1432). V. Quilis thanks to the Secretar\'{\i}a de Estado de Universidades,
Investigaci\'on y Desarrollo del Ministerio de Educaci\'on y
Cultura for a postdoctoral fellowship. Calculations were carried out in a 
SGI O2000 at the {\it Centre de Inform\'atica de la Universitat de Val\`encia}.
\end{acknowledgements}

\end{document}